\documentclass[onecolumn,amsmath,amssymb,12pt,superscriptaddress,nofootinbib]{revtex4}
\pdfoutput=1

\usepackage[latin1]{inputenc}
\usepackage[english]{babel}
\usepackage{amssymb}
\usepackage{amsmath}
\usepackage{amsthm}
\usepackage{float}
\usepackage[]{graphicx}
\usepackage[]{subfigure}
\usepackage{tensor}
\usepackage{color}
\usepackage{cancel}
\usepackage{setspace}
\usepackage{fancyhdr}
\usepackage{framed}
\usepackage{braket}
\usepackage{siunitx}
\usepackage{tikz}
\newcommand{\be}{\begin{equation}}
\newcommand{\ee}{\end{equation}}
\usepackage[bookmarks,linktocpage, colorlinks=true, plainpages = false, citecolor = blue,  linkcolor=blue, urlcolor = magenta, filecolor = blue]{hyperref} 

\usepackage{mathrsfs}
\usepackage[title]{appendix}
\numberwithin{equation}{section}

\begin{document}

\allowdisplaybreaks

\title{A simple superconducting\footnote{The published version comes with the word ``superconductor'' in the title as the more accurate word ``superconducting'' was not sent in time along with the proofs.} quantum interference device for testing gravity}
\vspace{.3in}

\author{Fay\c{c}al Hammad}
\email{fhammad@ubishops.ca}
\affiliation{Department of Physics and Astronomy, Bishop's University, 2600 College Street, Sherbrooke, QC, J1M~1Z7
Canada}
\affiliation{Physics Department, Champlain 
College-Lennoxville, 2580 College Street, Sherbrooke,  
QC, J1M~0C8 Canada}
\affiliation{D\'epartement de Physique, Universit\'e de Montr\'eal,\\
2900 Boulevard \'Edouard-Montpetit,
Montr\'eal, QC, H3T 1J4
Canada} 

\author{Alexandre Landry} \email{alexandre.landry.1@umontreal.ca} 
\affiliation{D\'epartement de Physique, Universit\'e de Montr\'eal,\\
2900 Boulevard \'Edouard-Montpetit,
Montr\'eal, QC, H3T 1J4
Canada}

\begin{abstract}
A simple tabletop setup based on a superconducting quantum interference device is proposed for testing the gravitational interaction. A D-shaped superconducting loop has the straight segment immersed inside a massive sphere while the half-circle segment is wrapped around the sphere. The superconducting condensate within the straight arm of the loop thus bathes inside a gravitational simple harmonic oscillator potential while the condensate in the half-circle arm bathes in the constant gravitational potential around the sphere. The resulting phase difference at the Josephson junctions on both sides of the straight arm induces a sinusoidal electric current that has a frequency determined by the precise gravitational potential due to the massive sphere.
\end{abstract}


\maketitle

\section{Introduction}
The smallness of the universal gravitational constant $G$ and the inverse square-law (ISL) --- the $1/r^2$-dependence of the gravitational interaction between any two massive objects separated by a distance $r$ --- are the main sources of difficulties one faces when attempting to test gravity and its universal constant $G$ to high precision. Any attempt to decrease the separation distance $r$ between the two masses in the hope of making the resulting gravitational force stronger and, hence, easier to be detected, is automatically accompanied by an increase in the strength of the other non-gravitational interactions, such as the van der Waals and Casimir forces to which rapidly add then the electric, the weak and strong nuclear forces that easily overwhelm the gravitational force as one keeps decreasing the separation distance $r$ \cite{Tests2003,LHC,Tests1985,Tests1993,TestsReview1999,Tests2001,Tests2002,Tests2007,Tests2009,SpaceISLTests,Tests2011,Tests2012,Tests2015,Tests2016,LISA,MICROSCOPE,ISLAND,Tests2019}. For this reason researchers have switched to indirect and more subtle tests based on quantum particles that do not rely on the gravitational force between a weakly separated pair of masses \cite{COWReview,UCNinEarth1,UCNinEarth2,UCNinEarth3,Biedermann,Kulin,Abele,CurvedSTInterferometry,Science}.

Among the more recent proposals in this direction by the present authors, are two setups based on two independent but somewhat related strategies. In Ref.\,\cite{COWBall}, the possibility of using quantum interference of cold neutrons (or any other neutral quantum particles for that matter) in a modified Colella-Overhauser-Werner (COW) experiment (see Ref.\,\cite{COWReview} for a review) was proposed. Such an experiment relies on the quantum interference induced by the gravitational field. It was shown in Ref.\,\cite{COWBall} that the gravitational potential felt by a quantum particle traveling inside a narrow cylindrical tunnel drilled along the diameter of a massive sphere is that of a one-dimensional harmonic oscillator. As such, it was shown that when making one of the two beams of an interferometer go through the tunnel while the other beam lies outside the mass source, a precise phase shift between the two beams of the interferometer results. Unfortunately, as such a strategy relies on a single particle interference, the induced phase shift calculated is extremely small and the setup is vulnerable to quantum decoherence. The expected phase shift is of the order $\sim10^{-2}\,$rad for a Newtonian potential and below $\sim10^{-4}\,$rad for any eventual deviation from the ISL.

In order for those small phase shifts to be detected one should manage to amplify the effect by either (i) increasing the mass and size of the gravitational source or (ii) by adding up such tiny phase shifts as they accumulate from a large number of affected particles. As the first possibility is rather expansive and would be accompanied by many technological challenges, one would rather opt for the second. Such a possibility is indeed feasible as shown in Ref.\,\cite{QHG}. The key idea \cite{QHG} is that the effects of gravity could be rendered macroscopic in mesoscopic systems \cite{Kiefer} by making many particles respond in unison to gravity as they do \cite{GravityLandauI,GravityLandauII} in the presence of a magnetic field in the quantum Hall effect \cite{QHEDiscovery}. In fact, it is found \cite{QHG} that the influence of gravity does show up in the quantum Hall effect but that such an effect would still not easily allow to distinguish between different possible deviations from the ISL. 

Sticking to this idea of exploiting quantum effects due to a large number of particles, we propose in this Letter to use quantum interference due to a large number of particles in a condensate state. In fact, it is well known that such a state, provided by superconductors, is a macroscopic state and, hence, any effect of gravity \cite{GravityJosephson,HolographyJunction} on it could show up at the macroscopic level as well. To achieve that, we propose here to use the so-called Josephson effect (see e.g., Ref.\,\cite{Josephson}). It is well known that such an effect is so sensitive that it allows one to measure minute magnetic fields down to strengths of the order of $10^{-15}\,$T. Similarly, the setup we propose here would allow to measure the gravitational interaction to very high precision. If at the first stages of its realization the setup does not reach the final desired precision to allow to test the ISL of gravity, it would still allow to measure the gravitational constant $G$ in a novel way and with high precision.

The remainder of this paper is structured as follows. In Sec.\,\ref{sec:TheSetup}, we present and describe the details of our proposed setup. In Sec.\,\ref{sec:TheAnalysis}, we conduct a rigorous mathematical analysis of the way the gravitational effect would manifest itself with the help of the setup. We provide a detailed derivation of the resulting gravitation-dependent frequency of the induced AC Josephson effect based on the ISL and various scenarios of deviations from the latter. In Sec.\,\ref{sec:Experimental}, we focus more on the eventual experimental realization of our setup, we provide its quantitative predictions and put into perspective its performance relative to those of the presently existing methods for testing gravity. We end this Letter with a brief conclusion and discussion section in which we highlight the main potentialities of our setup and possible future developments and improvements of the key idea behind the latter.

\section{The setup design}\label{sec:TheSetup}
\begin{figure}[H]
    \centering
    \includegraphics[scale=0.7]{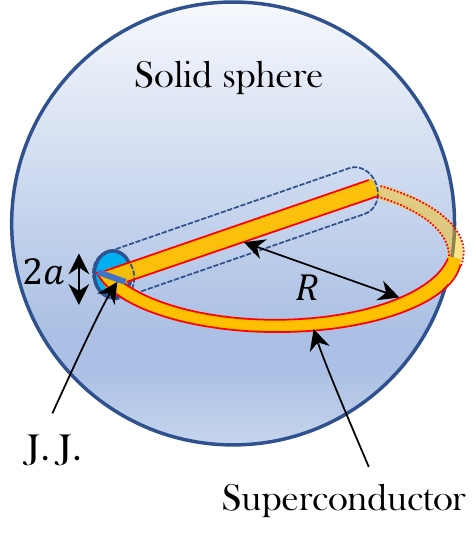}\qquad\qquad\qquad\includegraphics[scale=0.7]{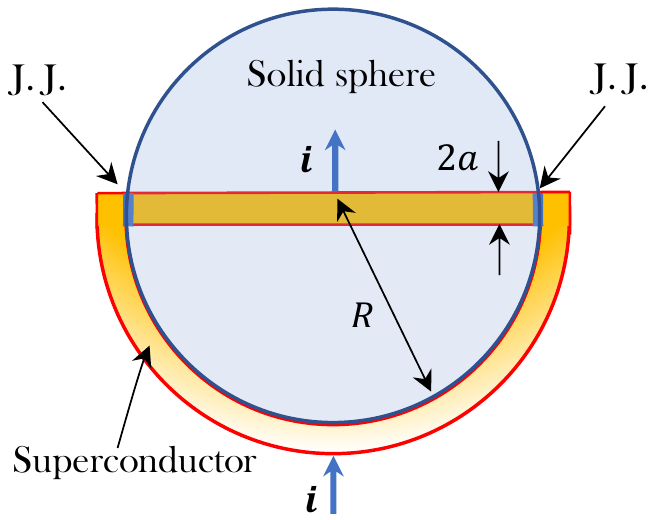}
    \caption{A D-shaped superconductor is partly immersed inside a massive sphere and partly wrapped around the latter. The left diagram gives a side view of the setup while the right one shows a top view. The current $i$ through the loop is due to the flow across the Josephson junctions (J.J.) on both sides of the straight segment of the superconductor.}
    \label{fig:Josephson}
\end{figure}

The setup we propose here is depicted in Figure\,\ref{fig:Josephson}. The setup consists of a massive solid sphere of radius $R$, inside of which is drilled a cylindrical tunnel of small radius $a$ along the diameter of the sphere. A D-shaped superconductor is used. The superconductor is thus made of two segments. The straight segment is connected through two superconductor-insulator-superconductor (S-I-S) Josephson junctions \cite{Josephson} to another segment curved into a half circle. The straight segment is immersed inside the sphere, along the drilled tunnel, while the curved segment is wrapped around the sphere. The flow of the electric current from one segment to the other is as shown in the top-view on the right in Figure\,{\ref{fig:Josephson}}.

Such a setup must then be cooled down to extremely low temperatures. Such a requirement, as we shall see in the next section, is what constrains the degree of sensitivity and precision of the device. In fact, in analogy to the low temperatures constraint imposed by any attempt to measure magnetic fields using a superconducting interference device (SQUID)\cite{Josephson}, the weakness of the gravitational interaction also requires very low temperatures to guarantee high performances of the setup. Nevertheless, we shall see that while the level of performance of the setup depends greatly on the low temperatures attainable, measuring the gravitational constant $G$ and testing gravity with such a device does not actually impose more constraints on the cryogenics than what other uses of SQUIDs do.

Very important also is the requirement to have a very uniform mass density for the solid sphere. This requirement, while not necessary for the gravitational effect detection proper, is going to render any correlation between an actual experimental measurement and the mathematical analysis we provide in the next section extremely rigorous. However, this requirement is not as experimentally challenging as is the case with the cryogenics. In fact, as we shall see in the next section, the size of the solid sphere does not have to be large. This is one of the great advantages of the present setup as compared to the proposal made in Ref.\cite{COWBall}. As such, attaining a uniform density for the sphere is much more easily achievable than the required low temperatures.

\section{Theory}\label{sec:TheAnalysis}
The principle behind our proposal relies on the remarkable difference between the form of the gravitational potential filling the interior of the tunnel inside the solid sphere and the form the gravitational potential has outside the sphere. The ultimate goal of our setup is thereby to use this difference in order to induce a current through the Josephson junctions connecting the two parts of the same superconductor spanning the inside and the outside regions of the solid sphere. In this section we are going to compute the frequency of the induced AC Josephson effect due to such a difference in the gravitational potential, first we investigate the case of a purely Newtonian potential and then we investigate two well-known possible cases for a deviation from the ISL: a Yukawa-like deviation and a power-law deviation.

\subsection{With a Newtonian potential}

As shown in great detail in Ref.~\cite{COWBall}, for a very small radius ($a\ll R$) of the tunnel inside the solid sphere, the Newtonian gravitational potential inside the latter at any position $x$ from the center of the sphere of radius $R$ and of mass density $\rho$ is given by, $V_g^N(x)=2\pi G\rho(R^2+\frac{1}{3}x^2)$. Here, we took the center of the sphere as the reference point for the gravitational potential. This gravitational potential is, however, nothing but the potential of a simple harmonic oscillator shifted by a constant term. Consequently, the superconducting condensate in the straight segment of the loop inside the sphere acquires the quantized energy of a simple harmonic oscillator of fundamental frequency $\omega_0=2\sqrt{\frac{\pi}{3}G\rho}$. Therefore, the energy of the Cooper pairs of effective mass $2m$ within the segment inside the sphere is given by,
\begin{equation}\label{EI(0)}
E_{I}=\mu+4\pi Gm\rho R^2+\hbar\omega_0\left(n+\frac{1}{2}\right),
\end{equation}
where $\mu$ is the chemical potential and $n$ is a non-negative integer. Meanwhile, the superconducting condensate in the half-circle segment outside the sphere acquires the Newtonian gravitational potential energy,
\begin{equation}\label{EOutNewtonian}
    E_O=\mu+2GMm/R=\mu+\frac{8\pi}{3} Gm\rho R^2
\end{equation}.

Thanks to the two Josephson junctions, the outside condensate is made to interact with the inside condensate via tunneling. Let us then denote by $K$ the real coupling amplitude between the outside and the inside condensates, and by $\ket{I}$ and $\ket{O}$ the inside and the outside states of the condensates, respectively. Then, the inside and outside Hamiltonians read $\mathcal{H}_I=E_I\ket{I}\bra{I}$ and $\mathcal{H}_O=E_O\ket{O}\bra{O}$, respectively, whereas the tunneling Hamiltonian reads $\mathcal{H}_T=K\left(\ket{I}\bra{O}+\ket{O}\bra{I}\right)$ \cite{Josephson}. The total system described by the state $\ket{\psi}=\psi_I\ket{I}+\psi_O\ket{O}$ then obeys the Schr\"odinger equation, $i\hbar\partial_t\ket{\psi}=(\mathcal{H}_I+\mathcal{H}_O+\mathcal{H}_T)\ket{\psi}$. Projecting this equation on the states $\ket{I}$ and $\ket{O}$, we arrive at the following coupled differential equations, respectively \cite{Josephson},
\begin{align}\label{CoupledSchrodinger}
i\hbar\frac{\partial\psi_I}{\partial t}&=E_I\psi_I+K\psi_O,\nonumber\\
i\hbar\frac{\partial\psi_O}{\partial t}&=E_O\psi_O+K\psi_I.
\end{align}
Let us now take the expression for the wave functions of the inside and outside condensates to be, respectively, $\psi_O=\sqrt{\rho_O}e^{i\phi_O}$ and $\psi_I=\sqrt{\rho_I}e^{i\phi_I}$, where $\rho_I$, $\rho_O$, $\phi_I$ and $\phi_O$ are all real. Inserting these expressions inside the equations (\ref{CoupledSchrodinger}), and separating the imaginary from the real parts, we find, after taking into account that $\partial_t\rho_I=-\partial_t\rho_O$ by virtue of charge conservation and $\rho_I=\rho_O=\rho_0$, the following two differential equations,
\begin{align}\label{JPhi}
\frac{\partial\rho_I}{\partial t}&=\frac{2K}{\hbar}\rho_0\sin\Delta\phi,\nonumber\\
\frac{\partial\Delta\phi}{\partial t}&=\frac{2\Delta E}{\hbar}.
\end{align}
Here, we have set $\Delta\phi=\phi_I-\phi_O$ and $\Delta E=E_I-E_O$. By integrating the second equation we find the expression for the induced phase difference across the junction to be given by, $\Delta\phi=2\hbar^{-1}\Delta E\,t+\phi_0$, for some arbitrary constant of integration $\phi_0$. Substituting this phase difference inside the first equation, and introducing the supercurrent density $J=\partial_t\rho_I$ of the pair condensate, we arrive at the following expression for the supercurrent density in terms of the energy $\Delta E$,
\begin{equation}\label{J}
J=J_0\sin\left(\phi_0+\frac{2\Delta E}{\hbar}t\right).
\end{equation}
Here, $J_0$ is called the critical supercurrent of the junction and the resulting effect is called an {\small{\sc AC}} Josephson effect \cite{Josephson}. It is this factor that requires low temperatures in order for any application of the {\small{\sc AC}} Josephson effect to be able to allow high precision measurements. In other words, the low temperatures constraint imposed on SQUID-based measurements is universal and arises regardless of what specific phenomena giving rise the energy difference $\Delta E$ one is trying to detect. We shall come back to this point in the last section. 

Inserting the gravitational energies $E_I$ and $E_O$ we found above for the inside and outside condensates, we deduce that the supercurrent density $J$ is a sinusoidal function of time, with a quantized frequency $\Omega_n$ given by,
\begin{equation}\label{OmegaNewton}
\Omega_n=\frac{8\pi Gm\rho R^2}{3\hbar}+\sqrt{\frac{16\pi G\rho}{3}}\left(n+\frac{1}{2}\right).
\end{equation}

The very interesting fact about this resulting frequency is that it is made of two parts. The second part is quantized and depends only on the gravitational constant $G$ and the density of the massive solid sphere $\rho$, whereas the first part is not quantized and depends, in addition, on the effective mass $2m$ of the Cooper pair and on the radius $R$ of the massive sphere. A closer inspection shows that if a massive sphere made of a metal as dense as platinum or osmium is used, the two terms would be of the same order only if the sphere has a radius of the order of a meter. This implies that the size of the sphere is not a constraint and that one can actually use a smaller sphere. For this reason, we may very well focus only in the Newtonian case on the second term of this formula and neglect the first term which becomes four orders of magnitude smaller than the second term for a $1$-cm radius massive sphere. However, for precision measurements purposes, one needs to include the first term as well. 

\subsection{With a Yukawa-like deviation from the ISL}
In this subsection we consider a deviation from the ISL for which the gravitational potential has the Yukawa-like form, $Gm_1m_2\frac{\alpha}{r}e^{-r/\lambda}$, where $r$ is the distance between the two point masses $m_1$ and $m_2$ and $\alpha$ is a dimensionless parameter. The latter parameter quantifies the strength of the deviation from the ISL and might in principle depend on the baryonic composition of the massive sphere, in which case the Weak Equivalence Principle (WEP) becomes violated. The parameter $\lambda$ has the dimensions of a length and represents the interaction range of the non-Newtonian gravitational force \cite{LHC}. With such an additional term to the Newtonian potential, the full potential inside the sphere found in Ref.\,\cite{COWBall} has a highly nonlinear form. Consequently, the Schr\"odinger equation cannot be solved analytically and the energy eigenvalues can only be extracted numerically, a task which lies beyond the scope of the present Letter. The best we can do here is seek an analytic estimate using perturbation theory with the main goal to demonstrate to full potential of our setup. Furthermore, as we shall see in the next section, the approximations we allow ourselves here are well justified by the very nature of our setup. 

For a very small radius of the tunnel, such that $a\ll R$ and $a^2\ll\alpha\lambda^2$, the gravitational potential inside the tunnel, found in Ref.~\cite{COWBall} at any distance $x$ from the center, can be approximated by, $V^N_g(x)+4\pi G\rho\alpha\lambda^2e^{-\frac{R}{\lambda}}\left[\frac{1}{x}(R+\lambda)\sinh\frac{x}{\lambda}-\cosh\frac{x}{\lambda}\right]$. Here, $V_g^N(x)$ represents the contribution we found above for the Newtonian part of the potential. In fact, the dominant contribution of these extra terms, as we shall see shortly, comes from an integration over $x$ around the value $x=R$ for which those extra terms become all proportional to $\alpha\lambda^2$. Treating such extra terms as a small perturbation then, the gravitational potential felt by the condensate inside the tunnel is that of a perturbed harmonic oscillator. Indeed, using the time-independent perturbation theory, we have the following first-order correction to the $n$-th energy eigenvalue (\ref{EI(0)}) of the unperturbed harmonic oscillator inside the tunnel:
\begin{align}\label{YukawaPerturbedIntegral}
E_{I}^{Y}&\approx8\pi Gm\rho\alpha\lambda^2 e^{-R/\lambda}\int_0^R\left[\frac{R+\lambda}{x}\sinh\left(\frac{x}{\lambda}\right)-\cosh\left(\frac{x}{\lambda}\right)\right]\psi_n^2(x)\,{\rm d}x.
\end{align}
Here the superscript $Y$ stands for Yukawa. We have introduced here the harmonic oscillator's normalized eingenfunctions \cite{BookQM} $\psi_n(x)=(2^nn!)^{-\frac{1}{2}}\left(\frac{b}{\pi}\right)^{\frac{1}{4}}e^{-\frac{b}{2}x^2}H_n(\sqrt{b}x)$, where $b=\frac{2m\omega_0}{\hbar}$ and $H_n(z)$ is the $n$-th Hermite polynomial \cite{BookFormulas}. Unfortunately, integral (\ref{YukawaPerturbedIntegral}) does not admit any simple analytic expression for general $n$. For this reason, we shall content ourselves in Sec.\,\ref{sec:Experimental} with performing a numerical evaluation of such an integral for specific values of the various parameters $R$, $\alpha$, $\lambda$ and $n$.

On the other hand, the extra gravitational energy the external condensate acquires due to the additional Yukawa-like term is \cite{COWBall},
\begin{equation}\label{EOutYukawa}
E_O^Y=8\pi Gm\rho\alpha\lambda^2e^{-\frac{R}{\lambda}}\left[\frac{R+\lambda}{R}\sinh\frac{R}{\lambda}-\cosh\frac{R}{\lambda}\right].
\end{equation}
The difference between these two energies adds to the energy difference $\Delta E$ between the inside and the outside condensates. The resulting new quantized frequency of the {\small{\sc AC}} Josephson effect then reads, 
\begin{equation}\label{OmegaYukawa}
    \Omega_n^{N+Y}\approx\frac{8\pi Gm\rho R^2}{3\hbar}+\sqrt{\frac{16\pi G\rho}{3}}\left(n+\frac{1}{2}\right)+\frac{2(E_O^Y-E_I^Y)}{\hbar}.
\end{equation}
The last two terms in this expression represent then a correction to the frequency induced by the purely Newtonian potential. Detecting such a correction in the measured frequency amounts then to detecting a Yukawa-like deviation from the ISL.


\subsection{With a power-law deviation from the ISL}
In this subsection we consider a deviation from the ISL that has the form of a power-law, $Gm_1m_2\frac{1}{r}\left(\frac{r_0}{r}\right)^q$, where $r_0$ is a parameter with the dimensions of length and it represents the interaction range of the non-Newtonian interaction. In addition, this parameter quantifies also the strength of the non-Newtonian interaction. It is model-dependent, and hence composition-dependent, which means that it might also in principle lead to violations of the WEP. The power $q$ is an integer \cite{LHC}.

With such an additional term in the Newtonian potential, with $q=1$ (for simplicity) and for a very small radius of the tunnel, such that $a\ll R$, the gravitational potential inside the tunnel, at any distance $x$ from the center, found in Ref.~\cite{COWBall}, can be approximated by, $V^N_g(x)+2\pi G\rho r_0\left(R+\frac{R^2-x^2}{2x}\ln\frac{R+x}{R-x}\right)$. Here, $V_g^N(x)$ represents again the contribution we found above for the Newtonian part of the potential. Treating the extra terms as a small perturbation, we find the following first-order correction to the $n$-th energy eigenvalue (\ref{EI(0)}) of the unperturbed harmonic oscillator inside the tunnel:
\begin{align}\label{PLPerturbedIntegral}
E_{I}^{PL}&\approx4\pi Gm\rho r_0\int_0^R\left(R+\frac{R^2-x^2}{2x}\ln\frac{R+x}{R-x}\right)\psi_n^2(x)\,{\rm d}x.
\end{align}
Here the superscript $PL$ stands for power-law. We have introduced here again the harmonic oscillator's normalized eingenfunctions $\psi_n(x)$. As with the case of the Yukawa-like deviation from the ISL, there is no simple analytic expression for integral (\ref{PLPerturbedIntegral}). We resort therefore to a numerical evaluation of such an integral by picking up specific values of the parameters $R$, $n$ and $r_0$. This is done in Sec.\,\ref{sec:Experimental} below. On the other hand, the extra gravitational energy the external condensate acquires due to the additional power-law term is \cite{COWBall},
\begin{equation}\label{EOutPL}
E_O^{PL}=4\pi Gm\rho Rr_0.
\end{equation}
The difference between these two energies adds to the energy difference $\Delta E$ between the inside and the outside condensates. The resulting new quantized frequency of the {\small{\sc AC}} Josephson effect then reads, 
\begin{equation}\label{OmegaPL}
    \Omega_n^{N+PL}\approx\frac{8\pi Gm\rho R^2}{3\hbar}+\sqrt{\frac{16\pi G\rho}{3}}\left(n+\frac{1}{2}\right)+\frac{2(E_O^{PL}-E_I^{PL})}{\hbar}.
\end{equation}
The last two terms in this expression represent again a correction to the frequency induced by the purely Newtonian potential. Detecting such a correction in the measured frequency amounts to detecting a power-law deviation from the ISL.

\section{Expected experimental results}\label{sec:Experimental}
As we have seen in the previous section, our setup does not only predict a quantized frequency for the {\small{\sc AC}} Josephson effect due to the gravitational effect of the solid sphere, but, more importantly, it predicts that the measured frequency would be different depending on whether or not gravity keeps obeying exactly the Newtonian ISL at short distances. 

\subsection{Measuring $G$}

Even the possibility of simply providing a new way for measuring the Newtonian gravitational attraction is in itself a remarkable feature of our setup. Indeed, the prevailing disagreement between the presently measured values of $G$ \cite{G} makes any new proposal for measuring the latter very attractive\cite{Tests2019}.   

In fact, the quantized frequency (\ref{OmegaNewton}) provided by our setup is directly determined by the gravitational constant $G$. Taking the latter to be the presently recommended value\cite{G} $G=6.67408\times10^{-11}\,$m$^3$/kgs$^2$, and using a $1\,$cm-radius solid sphere of, say, osmium of mass density $\rho=2.25872\times10^4\,$kg/m$^3$, we find the fundamental frequency $\omega_0=2.51288\times10^{-3}\,$Hz and the supercurrent frequency $\Omega_0$ evaluates to $2.52379\times10^{-3}\,$Hz for the ground state $n=0$ of the harmonic oscillator. This is an easily measurable frequency in the lab. The number of decimal places one allows inside the gravitational constant and inside the mass density of the sphere is what determines the precision one achieves for the frequency $\Omega_0$. As such, the setup could also be used the other way around. That is, by knowing the mass density $\rho$ of the solid sphere to a high precision and by measuring the frequency $\Omega_0$ of the supercurrent with an arbitrary number of decimal places, one would measure the gravitational constant $G$ up to the desired precision.

Now, one might argue that frequencies of the order of $10^{-3}\,$Hz are already too small to allow such large numbers of decimals in the measurements (see, however, Refs.\,\cite{LowFrequency, NanoHertz} for reports on the low-frequency applications of SQUIDs). Nonetheless, as we shall soon see below, there is an important beneficial experimental side-effect of the so-called gravitationally induced electric field, not yet taken into account here, which dramatically helps amplify the gravitational effect on the induced {\small{\sc AC}} Josephson current. Let us first, however, turn to the possibility of experimentally distinguishing between the purely Newtonian gravity and the two possible deviations from the ISL examined above. 
\subsection{Detecting a Yukawa-like deviation from the ISL}
To obtain the estimate (\ref{OmegaYukawa}) for the induced frequency, we had to make the assumption that $a\ll R$ in both cases and we had to make the extra assumption that $a^2\ll \alpha\lambda^2$ for the Yukawa-like deviation. These assumptions are actually not at all required by the formalism on which our proposal is based, but they do make the resulting gravitational potentials much easier to handle analytically when extracting the induced frequencies. In fact, all our setup requires is that one finds the gravitational potential energy difference between the condensates of the superconductor inside and outside of the solid sphere. As such, we might as well have relied on a numerical analysis when evaluating the gravitational potential at the outset. Although such an enterprise would have allowed a much wider applications range for our setup, it would nevertheless have taken us way beyond the goal and scope of the present Letter.

Fortunately, however, the very nature of our setup does indeed justify the approximations we adopted above in order to extract the analytic expressions \,(\ref{YukawaPerturbedIntegral}) and (\ref{PLPerturbedIntegral}). In fact, the D-shaped superconductor in our setup has, as is customary in standard SQUIDs \cite{DMBook,History}, a very small thickness which does not exceed a few micrometers allowing, thereby, a separation distance $2a$ between the two hemispheres to be of the order of a few micrometers as well or below. Therefore, even Yukawa-like deviations that are within a range such that, $\alpha\lambda^2\sim10^{-10}\,$m$^2$, would be largely within the framework of the analytic approximations we relied on above. 

It is worth noting here that there are presently many imposed constraints on the Yukawa parameters $\alpha$ and $\lambda$ coming from various experimental tests of gravity. These experiments include Earth-based ones\cite{Tests2003,LHC,Tests1985,Tests1993,TestsReview1999,Tests2001,Tests2002,Tests2007,Tests2009,Tests2011,Tests2012,Tests2015,Tests2016}, of which the E\"ot-Wash group\cite{Tests2002} has provided a neat excluded region on the Yukawa $(\alpha,\lambda)$-plane for $\alpha$ within the interval $[10^{-4}\sim10^{8}]$ and for $\lambda$ within the interval $[10^{-6}\,{\rm m}\sim10^{-2}\,{\rm m}]$. On the other hand, older\cite{TestsReview1999} and more recent\cite{SpaceISLTests,LISA,MICROSCOPE,ISLAND,Tests2019} space-based tests allow to put constraints on the Yukawa $(\alpha,\lambda)$-plane for $\alpha$ within the interval $[10^{-10}\sim10^{-1}]$ and for $\lambda$ within the interval $[10^{-2}\,{\rm m}\sim10^{14}\,{\rm m}]$. In all cases, however, the domain in the Yukawa $(\alpha,\lambda)$-plane for which $\alpha$ and $\lambda$ lie below certain regions still remains inaccessible. In fact, the major obstacle in reaching lower values of $\alpha$ and/or $\lambda$ in all experimental tests relying on the gravitational attraction between two masses --- hence, that are mechanical in nature --- remains of course the sensitivity of the setups to environmental noise. The sources of noise are mainly thermal fluctuations, seismic noise, and parasitic non-gravitational forces. In this respect, our setup stands out with its insensitivity to seismic vibrations and unwanted parasitic forces. As for thermal fluctuations, our setup reaches its full potentiality as soon as the required low temperatures for superconductivity to emerge are achieved. Thermal fluctuations around such temperatures, as we shall see below, do not alter the performance of the setup. With such advantages, our setup allows to effectively reach the presently inaccessible regions of the product $\alpha\lambda^2$.

For concreteness, we shall now resort to a numerical evaluation of the integral (\ref{YukawaPerturbedIntegral}) by adopting specific values of the various parameters of the experiment. If we assume that a $1\,$cm-radius massive sphere of osmium is used, then we have again $\omega_0=2.51288\times10^{-3}\,$Hz and thus $b=43.4124\,$m$^{-2}$. For the case $\alpha=10^{-4}$ and $\lambda=1\,$m, which still lies in the presently inaccessible region of the Yukawa $(\alpha,\lambda)$-plane, the correction $E_I^Y$ to the energy of the ground state of the harmonic oscillator is easily evaluated from Eq.\,(\ref{YukawaPerturbedIntegral}) to be $E_I^Y\approx\hbar\times1\times10^{-8}\,$Hz. On the other hand, with such parameters Eq.\,(\ref{EOutYukawa}) gives $E_O^Y\approx\hbar\times3.2\times10^{-7}\,$Hz. Therefore, the resulting correction to the measured frequency $\Omega_0$ as given by Eq.\,(\ref{OmegaYukawa}) evaluates to $6.2\times10^{-7}\,$Hz. For the case $\alpha=10^{8}$ and $\lambda=10^{-6}\,$m, which is also presently within the inaccessible region of the Yukawa $(\alpha,\lambda)$-plane, the correction $E_I^Y$ to the energy of the ground state of the harmonic oscillator is easily evaluated from Eq.\,(\ref{YukawaPerturbedIntegral}) to be $E_I^Y\approx\hbar\times10^{-14}\,$Hz. On the other hand, with such parameters Eq.\,(\ref{EOutYukawa}) gives $E_O^Y\approx\hbar\times2\times10^{-9}\,$Hz. Therefore, the resulting correction to the measured frequency $\Omega_0$ as given by Eq.\,(\ref{OmegaYukawa}) evaluates to $4\times10^{-9}\,$Hz.

\subsection{Detecting a power-law deviation from the ISL}
For a power-law deviation that has the parameter $r_0=10^{-3}\,$m, we find from Eq.\,(\ref{PLPerturbedIntegral}) that $E_I^{PL}\approx\hbar\times1.1\times10^{-7}\,$Hz whereas Eq.\,(\ref{EOutPL}) gives the value $E_O^{PL}\approx\hbar\times1.64\times10^{-6}\,$Hz. From Eq.\,(\ref{OmegaPL}) we find then that the correction to the frequency $\Omega_0$ is $3.06\times10^{-6}\,$Hz. Remarkably, however, our setup is, as mentioned above, immune to the other non-gravitational forces like the Casimir force and the van der Waals that usually plague any short-distance investigation of the gravitational attraction that rely on the standard mechanical devices. In fact, as the gravitational energy of the condensate outside the sphere as given in Eq.\,(\ref{EOutPL}) depends on the product $Rr_0$, we see that if we allow for a radius $R$ of the sphere of the order of $R=10\,$cm, the probed distance $r_0$ can go down to the order of $r_0=10^{-4}\,$m for the same frequency correction of the order of $1.40\times10^{-6}\,$Hz.

One might still argue that these frequencies are still way too small to allow any definitive conclusion as to the existence of non-Newtonian forces. However, as mentioned above, we shall look closer now at the effect of the induced electric field in the superconductor due to the gravitational field.

\subsection{The effect of the induced electric field}
Let us then revisit our results (\ref{OmegaNewton}), (\ref{OmegaYukawa}) and (\ref{OmegaPL}), and perform some ramifications on them. First of all, as it is shown in Ref.\,\cite{QHG}, by taking into account the well-known gravitationally induced electric field inside a metal under the influence of a gravitational field \cite{GravityInducedE1966,GravityInducedE1967,GravityInducedEThroughIons}, the effect of the gravitationally induced harmonic oscillator potential $V_g(x)$ on each of the electrons of the Cooper pairs is amplified by a factor of $\frac{1}{7}M/m$, where $M$ is the atomic mass of the superconductor's lattice. In fact, for a given gravitational potential $V_g(x)$ the induced electric potential felt by the electrons is $V_e(x)$ such that $eV_e(x)=\frac{1}{7}MV_g(x)$\cite{GravityInducedEThroughIons}. Thereby, the effective gravitational energy acquired by the Copper pairs is actually due to the induced electric potential energy and is given by $E=2m(\frac{1}{7}MV_g(x)/m)$.

Therefore, the frequency $\Omega_0$ due to the purely Newtonian potential in Eqs.\,(\ref{OmegaNewton}), (\ref{OmegaYukawa}) and (\ref{OmegaPL}) becomes amplified by the factor $\frac{1}{7}M/m$. Similarly, the correcting terms in Eqs.\,(\ref{OmegaYukawa}) and (\ref{OmegaPL}) become also multiplied by such a factor as it arises from Eqs.\,(\ref{YukawaPerturbedIntegral}), (\ref{EOutYukawa}), (\ref{PLPerturbedIntegral}) and (\ref{EOutPL}) giving the gravitational potential energies $E_I$ and $E_O$ of the condensate. For a superconductor made of lead, the atomic mass of which is $3.44064\times10^{-25}\,$kg, the amplifying factor $\frac{1}{7}M/m$ reaches the value $5.39575\times10^{4}$. Inserting such a factor into our estimates for the frequency $\Omega_0$, we find that the latter increases to $1.1723\,$Hz. Similarly, the corrections to the latter frequency become also affected. For the Yukawa-like deviation with $\alpha=10^{-4}$ and $\lambda=1\,$m the frequency correction is evaluated to be $0.0334\,$Hz. For $\alpha=10^8$ and $\lambda=10^{-6}\,$m, the frequency correction is $0.0002\,$Hz. For a power-law deviation, this amplifying factor allows to reach down to short distances of the order $r_0\sim10^{-4}\,$m even with a $10\,$cm-radius sphere of osmium, for which the frequency correction becomes as high as $0.07\,$Hz.


\section{Discussion and conclusion}
We have presented a new setup based on a superconductor made of two pieces. One piece is immersed inside a massive solid sphere and the other piece is wrapped around the latter. We saw that such a setup could be used not only to detect the usual Newtonian gravitational interaction and to measure the gravitational constant $G$ to very high precision, but also to detect any deviation from the ISL of gravity. In fact, any gravitational potential inside the tunnel that would be due to various possible deviations from the ISL introduced in the literature would modify the pure harmonic oscillator potential we derived for the inside of the solid sphere, and the Josephson frequency $\Omega_0$ would be slightly different from what is found from pure Newtonian gravity. Unfortunately, as the gravitational potential due to a deviation from the ISL is highly nonlinear \cite{COWBall}, a rigorous analytic expression for the gravitational potential energy of the condensate caused by such a deviation cannot be obtained. A systematic numerical analysis is required and will be attempted elsewhere. Nevertheless, the rough approximations we relied on here in order to evaluate such potentials were amply justified by the very nature of our setup which accommodates such estimates thanks to the possibility of using a very small diameter of the tunnel inside the sphere.

Yet, the mere capability of the setup to probe deviations from the ISL to unprecedentedly shorter distances offers new possibilities to use it to probe other gravity-related phenomena such dark matter and gravitational waves. In fact, in contrast to all the various ways SQUIDs have so far been proposed in the literature to test gravity \cite{History,DMBook,Mach,Gyros,DM2,Axion,DarkMatter}, our setup offers a much more compact design and much more economy. Now, while detecting gravitational waves using our setup is probably our of reach as gravitational wavelengths are of the order of thousands of kilometers, we believe that it would be possible to adapt the key principle behind our present proposal to achieve an accumulated effect in order to achieve such a detection. Nonetheless, it is not excluded that, thanks to its ability to detect deviations from the ISL, our setup could be used as a table-top experiment to detect dark matter. In fact, as it has already been suggested in Ref.\,\cite{DM&WEP}, dark matter could be responsible for the violation of the WEP. As such, detecting the latter would be equivalent to detecting the former. As we saw, our setup offers, indeed, the possibility of measuring the gravitational interaction to high precision using massive solid spheres made of different materials. Any slight deviation that would be due to different compositions of the used solid spheres would signal a violation of the WEP.

Let us now discuss the important issue of thermal fluctuations. In fact, one would expect that by exciting the harmonic oscillator within the sphere by using an applied voltage on both sides of the superconductor, one could benefit from having higher energy levels with larger integers $n$ which lead to an increase in the value of the quantized frequency and reduce, thereby, the effect of thermal noise. However, we should keep in mind that while the motion of the Cooper pair condensate is really governed by the potential of a simple harmonic oscillator inside the tunnel, the fact that the Josephson junctions constitute finite potential barriers on both sides alters the energy of the higher levels of the harmonic oscillator \cite{SHO}. Such barriers constitute indeed a real perturbation, albeit a tiny one due to the very high potential barrier made by the junctions compared to the gravitational potential. Only the low-lying states of the harmonic oscillator can thus be treated as pure unperturbed states to a very good approximation.

Although the experiment proposed here is supposed to be performed at extremely low temperatures in order to prevent thermal fluctuations from overwhelming the gravitational potential inside the tunnel, such fluctuations are inevitable as along as the experiment is conducted at finite temperatures. From the order of magnitude of the ground state of the harmonic oscillator one would wish indeed to ideally achieve a temperature of a few picokelvins for the thermal energy to be of the same order as that of the ground state. It is clear, however, that such extremely low temperatures are presently beyond reach in Earth-based laboratories in which the lowest temperatures achieved so far are tens of picokelvins and realized only on a few atoms \cite{PicoKelvins}. In superconductors, the temperatures achieved are of the order of a few millikelvins \cite{milliKelvins}. Yet, it is easy to check that the thermal fluctuations cannot screen the gravitational effect, and hence, that temperatures of the order of a picokelvin are not necessary for the present experiment to be successfully carried out based on the following classical argument \cite{Superconductivity}. Letting $\delta\phi$ denote the induced phase difference due to thermal fluctuations, expression (\ref{OmegaNewton}) yields $J=J_0\sin(\phi_0+\frac{2}{\hbar}Et+\delta\phi)$. As $\sin\delta\phi$ averages to zero, we have $\braket{J}=\bar{J}_0\sin(\phi_0+\frac{2}{\hbar}Et)$, where only the critical supercurrent becomes thus averaged to $\bar{J}_0=J_0\braket{\cos\delta\phi}$. At temperature $T$, the average critical supercurrent is then approximated by $e^{-k_BT/2K}$, where $k_B$ is the Boltzmann constant and $K$ is the coupling energy at the junction \cite{Superconductivity}. Therefore, the low temperature constraints are only dictated by the precision reached by the presently used current detectors. Presently used temperatures in SQUIDs of the order of a few kelvins or below are amply sufficient. 

Finally, we would like to conclude by emphasizing that the key idea behind our setup is to exploit the difference between the gravitational potentials inside and outside the solid sphere. We suggested here to use such a difference to create a phase difference between the condensates of two pieces of the same superconductor. Such a phase difference, in turn, manifests itself in the {\small{\sc AC}} Josephson effect. It is clear, then, that such a strategy can very well be carried out using other macroscopic quantum states than the one supplied by a superconductor. In particular, the superconductor in our setup can easily be replaced by a superfluid or any Bose-Einstein gas condensate, as long as the induced phase difference allows to give rise to a detectable interference effect in the manner reported in Refs.\,\cite{Superfluids,AtomSQUID}.

\section*{Acknowledgments}
The authors are grateful to the anonymous referees for their comments that helped improve the clarity and presentation of the manuscript. 
This work is supported by the Natural Sciences and Engineering Research
Council of Canada (NSERC) Discovery Grant (RGPIN-2017-05388).

\end{document}